# Optimal Pair Matching Combined with Machine Learning Predicts a Significant Reduction in Myocardial Infarction Risk in African Americans following Omega-3 Fatty Acid Supplementation

**Shudong Sun[1,2], Aki Hara[3], Laurel Johnstone[3], Brian Hallmark[4,5], Joseph C. Watkins[1,2], Cynthia A. Thomson[6], Susan M. Schembre[7], Susan Sergeant[8], Jason Umans[9], Guang Yao[10], Hao Helen Zhang[1,2], Floyd H. Chilton[3,4,5]\***

[1]Department of Mathematics, University of Arizona, Tucson, AZ, USA

[2]Statistics Interdisciplinary Program, University of Arizona, Tucson, AZ, USA

[3]School of Nutritional Sciences and Wellness, College of Agriculture and Life Sciences, University of Arizona, Tucson, AZ, USA

[4]BIO5 Institute, University of Arizona, Tucson, AZ, USA

[5]Center for Precision Nutrition and Wellness, University of Arizona, Tucson, AZ, USA

[6]Department of Health Promotion Sciences, Mel & Enid Zuckerman College of Public Health, University of Arizona, Tucson, AZ, USA

[7]Lombardi Comprehensive Cancer Center, Georgetown University, Washington, DC, USA

[8]Department of Biochemistry, Wake Forest School of Medicine, Winston-Salem, NC, USA

[9]Department of Medicine, Georgetown University Medical Center, Washington, DC, USA

[10]Department of Molecular and Cellular Biology, University of Arizona, Tucson, AZ, USA

**\*Correspondence:** Floyd H. Chilton fchilton@arizona.edu

**Abstract**

Conflicting clinical trial results on omega-3 highly unsaturated fatty acids (n-3 HUFA) have prompted uncertainty about their cardioprotective effects. While the VITAL trial found no overall cardiovascular benefit from n-3 HUFA supplementation, its substantial African American (AfAm) enrollment provided a unique opportunity to explore racial differences in response to n-3 HUFA supplementation. The current observational study aimed to simulate randomized clinical trial (RCT) conditions by matching 3,766 AfAm and 15,553 non-Hispanic White (NHW) individuals from the VITAL trial utilizing propensity score matching to address the limitations related to differences in confounding variables between the two groups. Within matched groups (3,766 AfAm and 3,766 NHW), n-3 HUFA supplementation's impact on myocardial infarction (MI), stroke, and cardiovascular disease (CVD) mortality was assessed. A weighted decision tree analysis revealed belonging to the n-3 supplementation group as the most significant predictor of MI among AfAm but not NHW. Further logistic regression using the LASSO method and bootstrap estimation of standard errors indicated n-3 supplementation significantly lowered MI risk in AfAm (OR 0.17, 95% CI [0.048, 0.60]), with no such effect in NHW. This study underscores the critical need for future RCT to explore racial disparities in MI risk associated with n-3 HUFA supplementation and highlights potential causal differences between supplementation health outcomes in AfAm versus NHW populations.

**Introduction**

In 1999, the landmark GISSI-Prevenzione (GISSI-P) randomized, placebo-controlled trial (RCT) was published reporting the effects of daily supplementation with 1 g/day of n-3 highly unsaturated fatty acids (n-3 HUFAs), eicosapentaenoic acid (EPA) + docosahexaenoic acid (DHA), in patients (N=2,836 n-3 HUFA vs 2,830 placebo) who had recently experienced a myocardial infarction (MI) [1]. Over the 4-year study, researchers reported a 14% risk reduction in the combined primary endpoint of relative risk and an overall 20% reduction in fatal cardiovascular events. This groundbreaking study, however, was met with mixed results in subsequent studies, which questioned the additive benefits of n-3 HUFA supplementation to modern cardiovascular therapies, particularly at lower doses [2,3]. Further investigations, including a 2018 comprehensive systematic review, containing 79 RCTs with 112,059 patients found an overall 8% reduction in cardiac death associated with n-3 HUFA supplementation [4].

Yet, the landscape of n-3 HUFA research continues to be marked by variability, as seen in the differing outcomes of notable recent large RCTs such as ASCEND, VITAL and REDUCE-IT. ACSEND examined the impact of n-3 HUFA supplementation on individuals with diabetes, and found no significant differences in the risk of serious vascular events between n-3 HUFA supplementation and placebo groups [5]. VITAL tested the impact of n-3 HUFA and vitamin D3 supplementation on primary prevention of cardiovascular disease (CVD) and cancer, again finding no difference between the n-3 HUFA intervention and placebo groups [6]. In contrast, the REDUCE-IT clinical trial provided higher doses of EPA as an ethyl ester to patients at high risk of CVD and demonstrated a 25% reduction in composite CVD morbidity and mortality (including CVD mortality, non-fatal MI, non-fatal stroke, cardiovascular revascularization or unstable angina) in the treatment group [7]. Subsequent meta-analyses and reviews suggest discordant clinical trial results are likely related to numerous factors, including different doses and forms of n-3 HUFAs provided, background dietary levels of n-3 HUFAs, CVD and diabetes-related risk factors, and medications, each of which may limit the measurable effects of n-3 HUFAs [8-11].

The RCTs to date have generally demonstrated a lack of diversity across the study populations, implicitly suggesting a one-size-fits-all response to n-3 HUFA supplementation across racial groups. Yet, arachidonic acid (ARA) production and metabolism, a major driver of CVD risk [12], varies widely across racial groups related to differences in the frequency of key genetic variants particularly within the fatty acid desaturase (*FADS*) gene cluster that influence the biosynthesis of HUFAs. Importantly, compared to NHW, AfAm exhibit significantly higher frequencies of fatty acid desaturase (*FADS)* genetic variants which are linked to a more efficient conversion of the dietary omega-6 (n-6) polyunsaturated fatty acid (PUFA) linoleic acid (LA) to ARA [13,14]. These genetic factors and other race-associated variation [13] coupled with the high dietary levels (6-8% total energy intake) of LA in Western diets (MWD) have the potential to contribute to pronounced differences in the balance of ARA-derived pro-inflammatory and pro-thrombotic metabolites relative to EPA and DHA-based anti-inflammatory, anti-thrombotic, and pro-resolution oxylipin profiles among AfAm compared to NHW populations [15].

Leveraging the diversity of the VITAL trial and its substantial AfAm enrollment, follow-up analyses suggest a notable reduction in MI risk with n-3 HUFA supplementation in AfAm, depending on baseline cardiovascular risk and fish consumption [16-18]. However, disparities in key covariates such as age, BMI, and comorbidities across racial and ethnic groups within VITAL raise questions as to whether factors other than race may influence whether n-3 HUFAs are effective. Furthermore, an analysis of VITAL using hierarchical composite CVD outcomes based on win ratio demonstrated that participants with low fish consumption at baseline benefited more than those with high consumption when randomized to receive n-3 HUFA supplementation [19]. Importantly, another recent secondary analysis of an RCT demonstrated significant racial differences between AfAm and NHW mothers in baseline levels of DHA and the efficacy of DHA supplementation in preventing preterm birth [20].

The goal of the current study was to employ propensity score matching to address limitations of secondary data analysis related to differences in confounding variables between AfAm and NHW. This study retrospectively analyzed the potential impact of n-3 HUFA supplementation compared with a placebo on the risk of MI in a heterogeneous population, stratified into two ethnic groups. Specifically, we paired NHW participants with an AfAm subgroup, based on their covariate profiles and then leveraged machine learning methods to carefully curate a dataset of 3,766 AfAm and matched NHW individuals. Using this methodology, we were able to simulate RCT conditions within observational data to assess risk of MI, stroke and CVD mortality with n-3 HUFA supplementation in the AfAm group as compared to NHW. Findings from this study provide the best evidence to date that highlight the potential of n-3 HUFA supplementation to prevent MI in AfAm individuals. They also emphasize the necessity for considering race in the design and power calculations of future trials utilizing n-3 HUFA supplementation.

**Materials & Methods**

**Original Study Design.** In the original VITAL study [6], a randomized, double-blind, placebo-controlled trial with a two-by-two factorial design, the researchers assessed the benefits and risks of supplementation with vitamin D3 (2000 IU per day) and n−3 HUFA (1 g per day, provided as a fish-oil capsule containing 840 mg of n−3 HUFA, including 460 mg of eicosapentaenoic acid [EPA] and 380 mg of docosahexaenoic acid [DHA]) for the primary prevention of cardiovascular disease and cancer in men aged 50 and older and women aged 55 and older in the United States. Additionally, AfAm participants were oversampled (5,107 participants) to facilitate a focused evaluation of the effects in this demographic. Participants were eligible if they had no history of cancer (except non-melanoma skin cancer), heart attack, stroke, transient ischemic attack, or coronary revascularization, with no other cancer or vascular risk factors considered for selection [6].

Certain safety criteria were applied to exclude individuals with renal failure or on dialysis, high or low calcium levels, parathyroid disorders, severe liver disease (cirrhosis), or chronic conditions associated with increased hypercalcemia risk (such as sarcoidosis, Wegener's granulomatosis, or chronic tuberculosis). Other exclusions included those on anti-coagulant medications, allergies to soy (in the vitamin D placebo) or fish, and serious illnesses that could impede participation [6,18].

Participants in the VITAL trial were randomized to receive either n−3 HUFA, vitamin D, both active agents, or both placebos between November 2011 and March 2014. The intervention period concluded on December 31, 2017, providing a median follow-up duration of 5.3 years, with a range from 3.8 to 6.1 years [6]. Data collection involved baseline questionnaires capturing clinical and lifestyle risk factors, dietary assessments, and blood samples for measuring plasma Omega-3 Index. Annual questionnaires monitored adherence to supplementation, side effects, and the development of major illnesses or updates in risk factors. The primary and secondary endpoints included major cardiovascular events, various cancers, and associated mortality [6,18]. The study was registered on ClinicalTrials.gov with the identifier NCT01169259.

**Data Preprocessing**. Data used for the reanalyses were obtained from the public repository of the original VITAL trial at Project Data Sphere (https://data.projectdatasphere.org/projectdatasphere/html/content/454). In our study, the initial dataset comprised data from 25,871 participants, which included NHW individuals (18,046; 71%), AfAm individuals (5,106; 20%), and individuals from other races (2,719; 9%). From this initial pool, a subset of data was excluded due to race categorization outside the primary focus groups (2,719 participants) and incomplete records (3,833 participants), leaving 19,319 participants for preprocessing. This preprocessing involved cleaning, standardizing, and preparing the data for analysis, yielding a refined dataset with 15,553 NHW and 3,766 AfAm participants.

To address potential confounding factors and ensure comparability between groups, a matching algorithm was applied (see below for details). Non-matched NHW individuals (11,787) were excluded from the subsequent analysis to balance the racial composition of the study groups, resulting in an equal number of NHW and AfAm participants (3,766 in each group). The final dataset, therefore, consisted of 7,532 individuals, divided equally between the two racial categories. This meticulous approach in data handling aimed to minimize bias and facilitate a fair comparative analysis of outcomes across the racial groups included in the study. For a visual representation of the data processing and matching procedure, see **Figure 1**.

**Figure 1**: Flowchart of data processing

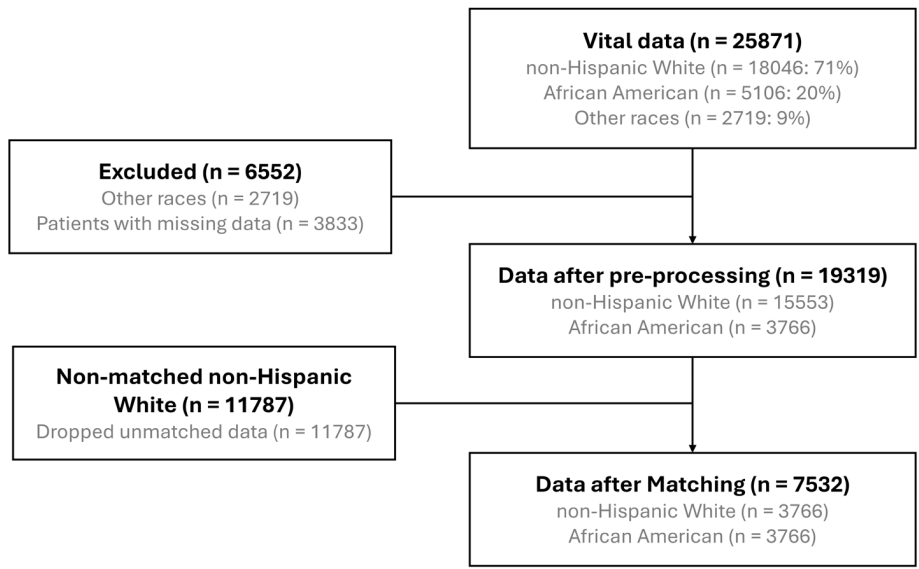

**Propensity Score Matching.** Propensity score matching [21,22] was employed to select a matched NHW individual for each AfAm individual in the dataset. All variables collected in the original research, except for two that were excluded due to a large proportion of missing values (see **Table S1**), were used for propensity score matching. Individuals with missing values in these remaining features were removed. The R package *MatchIt* [23] was utilized to implement three different matching methods [24-26]: optimal matching, nearest neighbor matching, and genetic matching. The first two methods are based on estimating the propensity score using logistic regression. Nearest neighbor matching, also known as greedy matching, involves sequentially pairing treated units with the closest eligible control unit. Optimal pair matching selects matches that collectively optimize an overall criterion. In this study, the criterion was the sum of absolute pair distances in the matched sample. The third method, genetic matching, employed a genetic algorithm, which optimized non-differentiable objective functions, to determine scaling factors for each variable in a generalized Mahalanobis distance formula. The algorithm seeks to achieve covariate balance. After obtaining the scaling factors, nearest neighbor matching was performed using the scaled generalized Mahalanobis distance. To evaluate the performance of different matching methods, the standardized mean difference (SMD) was evaluated, and optimal pair matching outperformed the other approaches and was thereby utilized.

**Survival Analysis and Kaplan-Meier Curves**. Subgroup analysis was performed using the participants chosen by optimal pair matching for the outcome of MI, stratified by race. Only NHW and AfAm subjects selected by the optimal pairs matching algorithm were included. Statistical calculations were performed in R (version 4.2.2) using the survival [27] (version 3.4-0) and survminer (https://rpkgs.datanovia.com/survminer/index.html, version 0.4.9) packages. Kaplan-Meier survival curves were produced separately for each racial group, and p-values for the effect of the n-3 HUFA treatment on the occurrence of MI were calculated using the log-rank test.

**Logistic Regression Analysis with LASSO for Variable Selection and Bootstrap for Standard Error Estimation**. For this analysis, both racial subgroups were included in one dataset with race encoded as a variable along with all other variables retained after filtering for missing data (see Data Preprocessing above). Interaction terms between n-3

HUFA supplementation and all other variables were introduced to identify potential interactions with n-3 HUFA supplementation. To classify patients with MI (or stroke or cardiovascular death) versus those without MI, a weighted LASSO regression model [28] was built using the *glmnet* [29] package in R for variable selection. A 5-fold cross-validation method was employed to tune the LASSO model. Subsequently, a logistic regression model was fitted considering only the variables selected by the LASSO model. To estimate the standard errors, both non-parametric bootstrap and parametric bootstrap methods [30] were employed. Using the logistic regression coefficients and the standard errors estimated through bootstrapping, p-values, odds ratios, and corresponding 95% confidence intervals were calculated. During the bootstrap procedure, models with large distances for critical coefficients were excluded to avoid drawing poor samples.

**Decision Tree**. The matched data were utilized as input to build a predictive model, specifically a decision tree, using the classification and regression trees (CART) algorithm [31] implemented in the R package *rpart* [32]. The AfAm and NHW subgroups were independently utilized to construct the trees, facilitating a comparison of the outcomes. Since the dataset was severely unbalanced, a weighted tree algorithm was applied with the weight ratio equal to the number of individuals with MI (or stroke or cardiovascular death) to the number of individuals without these outcomes. The Gini index was employed to determine the tree split points, and a minimum node size of 20 was set as a requirement. To optimize the tree model and assess its prediction accuracy, a 5-fold cross-validation method was employed. To prevent overfitting, the tree was pruned using the 1-SE rule. This rule identifies the number of splits that yield the smallest cross-validation error, adds the corresponding standard error, and selects the fewest splits where the cross-validation error remains smaller than this combined value.

**Results**

**Utilization of Optimal Pair Matching to Balance Potential Confounding Variables**

To reduce bias arising from confounding variables in the original VITAL trial dataset, propensity score-based matching was employed to create a curated set of matched samples from the two populations for comparison. After initial testing of three different matching procedures, the optimal pair matching algorithm was selected due to its superiority in minimizing the standardized mean difference between matched pairs (see Methods).

Following data preprocessing and matching, 7532 participants (3766 each from the AfAm NHW populations) remained (see **Figure 1**) with complete data for 19 study variables (**Table S1**). As noted earlier, the original VITAL study did not stratify based on race to compare these groups, and many covariates had markedly different distributions across racial groups (**Table S2**). For example, the average age for NHW was 67.4 years vs 62.4 years for AfAm. The smoking rate was 14% in NHW vs 5% in AfAm, and the diabetes comorbidity was 23.1% in NHW and 10% in AfAm. The characteristics of the pared-down cohort after optimal pair matching are detailed in **Table 1**. Important variables such as age, smoking, medications, fish consumption, and cardiovascular risk factors were comparable between AfAm and NHW participants in this carefully matched dataset.

**Table 1**: Demographic Table after Optimal Pairs Matching.

| Characteristic | Total (N=7532) | AfAm (N=3766) | NHW (N=3766) |
|---|---|---|---|
| Age | 62.8±6.3 | 62.4±6.6 | 63.1±5.9 |
| Female | 4587 (60.9) | 2347 (62.3) | 2240 (59.5) |

|  |  |  |  |  |
|---|---|---|---|---|
| BMI |  | 30.3±6.6 | 30.6±6.5 | 29.9±6.1 |
| Current smoking |  | 979 (13.0) | 528 (14.0) | 451 (12.0) |
| Hypertension medication |  | 4754 (63.1) | 2446 (64.9) | 2308 (61.3) |
| Cholesterol medication |  | 2451 (32.5) | 1200 (31.9) | 1251 (33.2) |
| Statin use |  | 2279 (30.3) | 1114 (29.6) | 1165 (30.9) |
| Diabetes |  | 1617 (21.5) | 871 (23.1) | 746 (19.8) |
| Diabetes medication |  | 1267 (16.8) | 675 (17.9) | 592 (15.7) |
| Parental history of MI |  | 1161 (15.4) | 591 (15.7) | 570 (15.1) |
| Fish consumption (≥1.5/wk) |  | 3750 (49.8) | 1890 (50.2) | 1860 (49.4) |
| Aspirin use |  | 2885 (38.3) | 1431 (38.0) | 1454 (38.6) |
| Vitamin D supplements |  | 2271 (30.2) | 1096 (29.1) | 1175 (31.2) |
| CVD risk factors | 0 | 1661 (22.1) | 778 (20.7) | 883 (23.4) |
|  | 1 | 2634 (35.0) | 1315 (34.9) | 1319 (35.0) |
|  | >1 | 3237 (43.0) | 1673 (44.4) | 1564 (41.5) |

**Effect of n-3 HUFA Supplementation on MI in AfAm and NHW Participants**

Following this matching approach, Kaplan-Meier analysis was initially used to compare rates of MI between groups receiving n-3 HUFA supplementation and those given placebo, stratified by race. **Figure 2** illustrates the resulting survival curves and MI events over the six-year study. AfAm participants who were randomized to n-3 HUFA supplementation showed a marked reduction in MI compared to those who received placebo (p=0.00034), indicating a statistically significant effect in this subgroup. Conversely, no significant reduction in MI was observed in NHW participants (p=0.59).

**Figure 2**: Kaplan-Meier Survival Curves for Myocardial Infarction in n-3 HUFA Treatment and Placebo Groups, Faceted by Race/Ancestry, for Participants Selected by Optimal Pair Matching. a. Kaplan-Meier plots and the number of subjects at each event point for NHW participants matched with AfAm participants. b. Kaplan-Meier plots and the number

of subjects at each event point for AfAm participants. n-3 HUFA, n-3 HUFA supplementation group; placebo, placebo group. Due to the low incidence of MI in the sampled populations, the curves are presented starting from a y-axis value of 0.97 to enable visualization of the differences. The tables below the plots indicate the number of subjects in each group still in the study and MI-free at each six-month time point. Unadjusted p-values for the treatment effect were calculated for each racial group using the log-rank test.

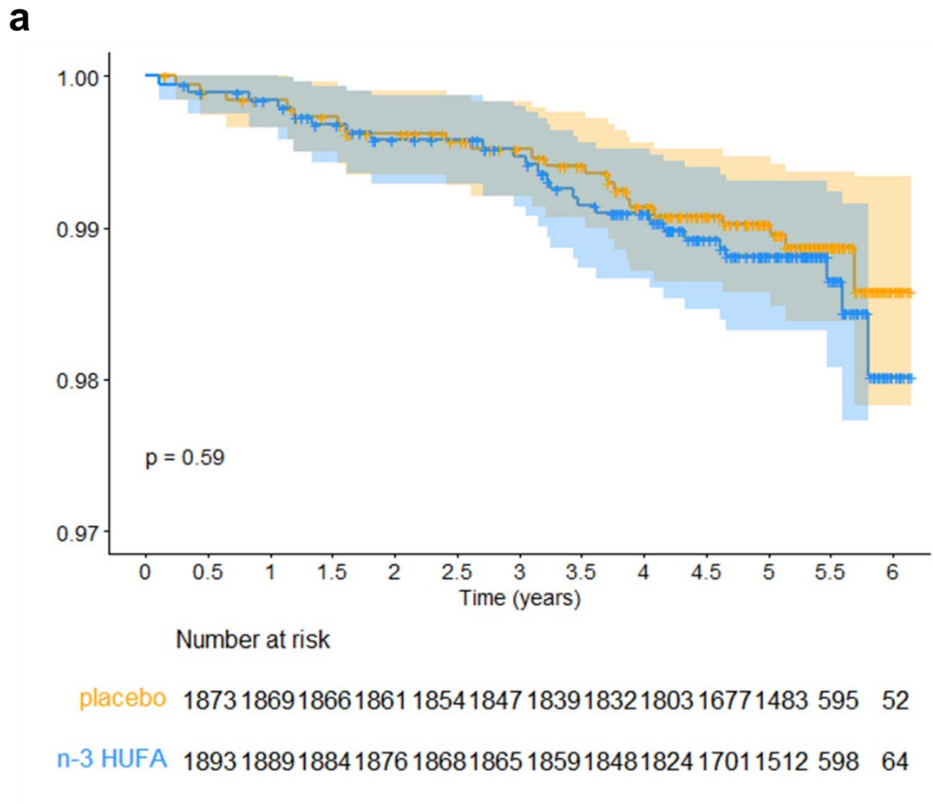

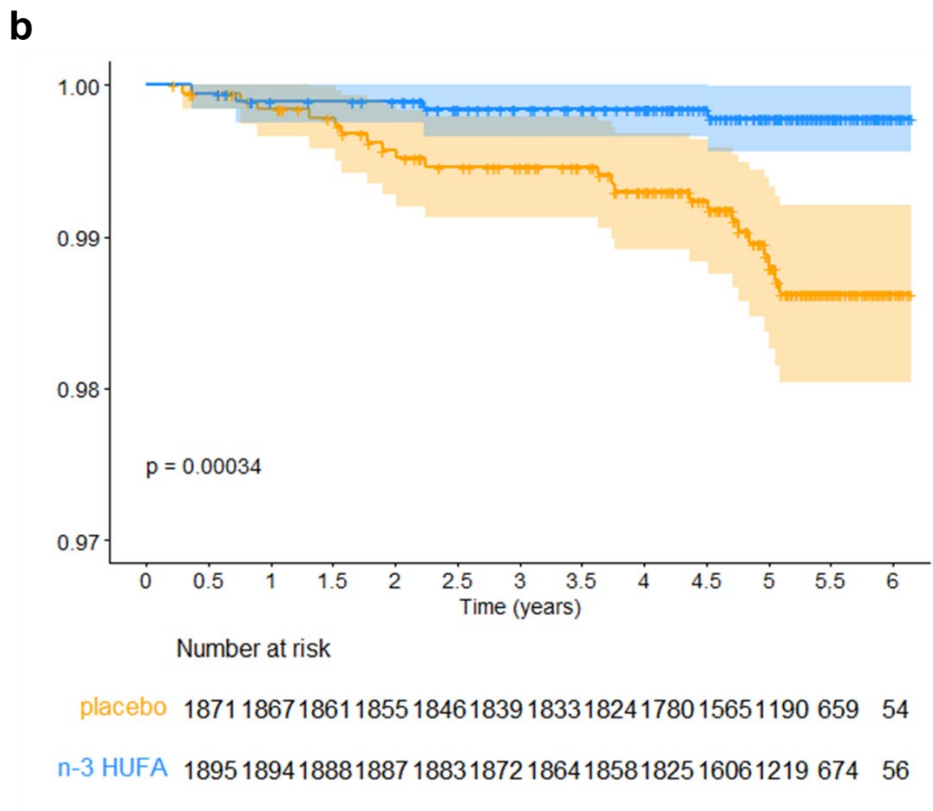

## Logistic Regression Analysis with LASSO to Select Important Variables and Bootstrap to Estimate the Standard Errors

Logistic regression analysis with interaction terms between n-3 HUFA supplementation and other variables (including race), employing the Least Absolute Shrinkage and Selection Operator (LASSO) for variable selection and bootstrap methods for estimating standard errors, identified n-3 HUFA supplementation as the most significant predictor of MI incidence among AfAm participants (**Table 2**). These series of analyses were performed on a combined population of matched AfAm and NHW, and race was factored in as a variable. The odds ratio (OR) for MI incidence in AfAm with n-3 HUFA supplementation compared to placebo was 0.17, with a 95% confidence interval (CI) ranging from 0.048 to 0.60. Conversely, for NHW participants, the analysis indicated no significant effect of n-3 HUFA supplementation on MI risk, with an OR of 1.0 and a 95% CI of [0.74, 1.34]. These findings remained consistent across both non-parametric and parametric bootstrap analyses (**Table 2**).

**Table 2:** Regression Results for MI Outcomes Using LASSO-Selected Variables with Both AfAm and NHW Participants After Optimal Pairs Matching. Includes interaction terms. Std. Errors, p-values and 95% CI for OR are calculated using non-parametric bootstrap and parametric bootstrap. OR, odds ratio; CI, confidence interval. *n-3 HUFA supplementation OR (95% CI) for AfAm and NHW subgroups.

| Variables | Estimate | OR | Non-parametric Bootstrap | | | Parametric Bootstrap | | |
|---|---|---|---|---|---|---|---|---|
| | | | Std. Error | P-value | 95% CI for OR | Std. Error | P-value | 95% CI for OR |
| (Intercept) | -3.7599 | 0.0233 | 1.9965 | 0.0597 | (0.0005, 1.1656) | 1.8195 | 0.0388 | (0.0007, 0.8239) |
| Female | -0.7184 | 0.4875 | 0.3501 | 0.0402 | (0.2455, 0.9683) | 0.3448 | 0.0372 | (0.2480, 0.9583) |
| Age | 0.0552 | 1.0568 | 0.0277 | 0.0463 | (1.0009, 1.1157) | 0.0251 | 0.0279 | (1.0060, 1.1100) |
| BMI | 0.0304 | 1.0309 | 0.0249 | 0.2221 | (0.9818, 1.0824) | 0.0234 | 0.1939 | (0.9847, 1.0792) |
| Current Smoker | 0.7167 | 2.0477 | 0.4155 | 0.0845 | (0.9069, 4.6232) | 0.4471 | 0.1089 | (0.8525, 4.9186) |
| Diabetes | 0.4599 | 1.5839 | 0.4255 | 0.2798 | (0.6879, 3.6469) | 0.3381 | 0.1738 | (0.8165, 3.0728) |
| Fish consumption (1.5/wk) | -0.7424 | 0.4760 | 0.3755 | 0.0480 | (0.2280, 0.9936) | 0.3274 | 0.0234 | (0.2505, 0.9042) |
| n-3 HUFA supplementation x AfAm | -1.7747 | 0.1695 | 0.6410 | 0.0056 | (0.0483, 0.5955)* | 0.6944 | 0.0106 | (0.0435, 0.6613)* |
| n-3 HUFA supplementation x NHW | 0.0000 | 1.0000 | 0.1508 | 1.0000 | (0.7442, 1.3438)* | 0.1566 | 1.0000 | (0.7357, 1.3593)* |

**Weighted Decision Tree Analysis of MI in AfAm and NHW Participants**

Weighted decision trees were constructed to identify the most predictive variables of MI incidence among AfAm and NHW group, analyzed separately. **Figure 3a** presents the decision tree for AfAm participants, where the first branching point underscores that assignment to the n-3 HUFA supplementation group as the predominant factor for MI prevention in this subgroup. Subsequent branches identify BMI, age, and the presence of diabetes as significant variables influencing MI incidence among AfAm participants. Conversely, the decision tree for NHW participants, shown in **Figure 3b**, illustrates that n-3 HUFA supplementation did not significantly affect MI incidence. Instead, BMI and sex emerged as primary variables associated with MI outcomes.

**Figure 3:** Weighted Decision Tree for Predicting the Incidence of Myocardial Infarction. a. Decision Tree for AfAm participants after optimal pair matching, with a total of n=3,766. The weight ratio of non-diseased to diseased participants is 3,740:26, equating to 144:1. b. Decision Tree for NHW participants after optimal pair matching, with a total of n=3,766. The weight ratio of non-diseased to diseased participants is 3,720:46, which corresponds to 81:1. n-3 HUFA supplementation, active or placebo; BMI, body mass index at randomization, kg/m2; age, age at randomization to VITAL study, years; diabetes, baseline diabetes, yes or no; sex, female or male.

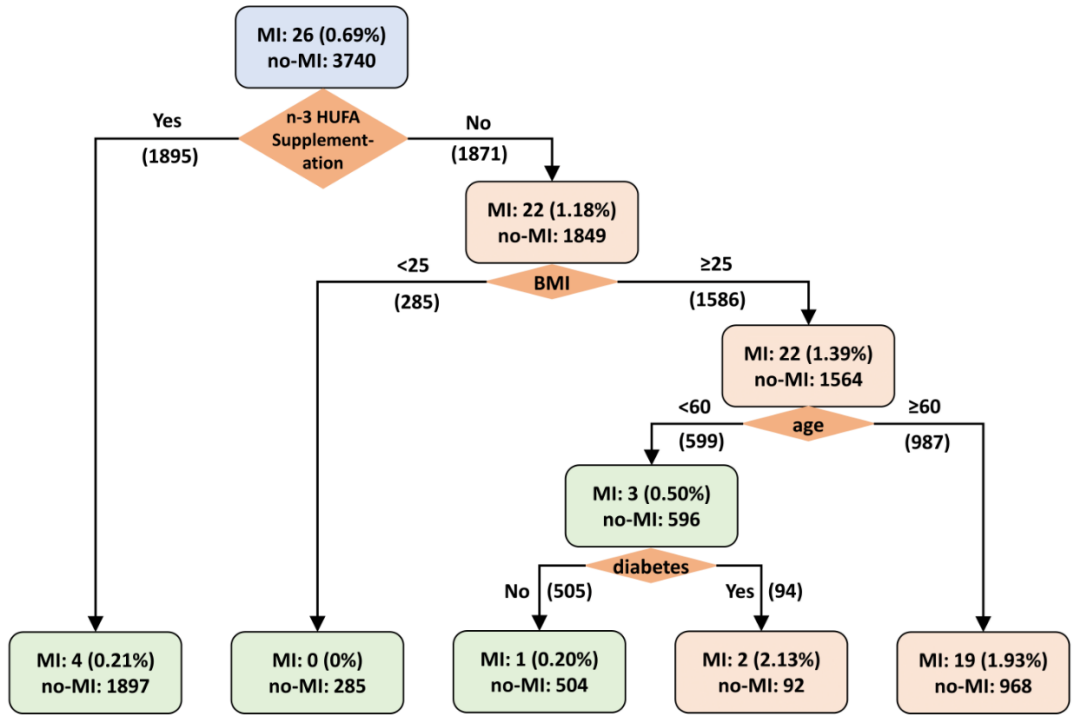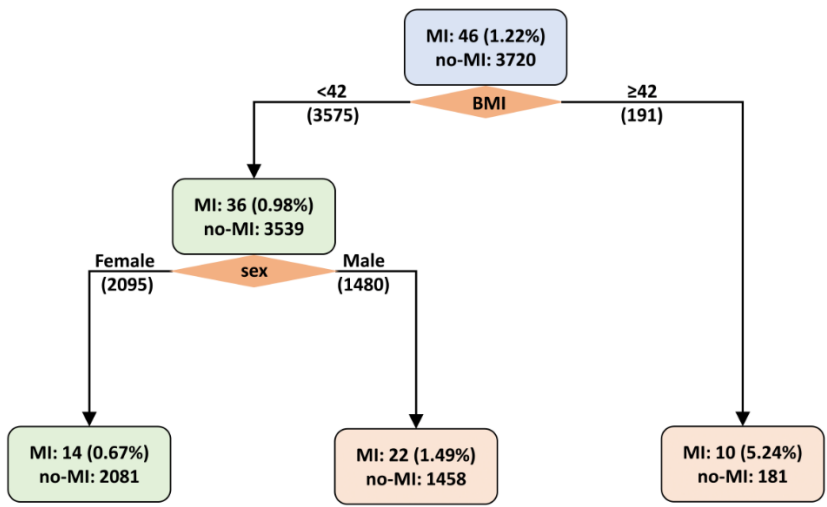

**LASSO Regression Analysis and Weighted Decision Tree of Stroke or CVD Mortality in AfAm and NHW Participants**

Neither the decision tree nor the logistic regression with LASSO analyses identified randomization to n-3 HUFA supplementation as an important factor that impacts stroke or CVD mortality in AfAm or NHW groups (**Tables S3 & S4, Figures S1 & S2).** In the combined population logistic regression model addressing stroke as the outcome, only age remained a significant factor. Although assignment to the n-3 HUFA group was among the variables selected by the LASSO algorithm, its interaction with race did not prove to be significant for either the AfAm or NHW subgroups (**Table S3**). Similarly, in the regression analysis focusing on CVD mortality, age and current smoking were the only significant predictors, and n-3 HUFA supplementation did not appear among the LASSO-selected variables (**Table S4**). Both regression models were further validated using parametric bootstrap analysis, yielding results consistent with those obtained from the non-parametric bootstrap approach (**Tables S3 & S4**). The decision tree for stroke identified age and BMI as predictive factors for both AfAm and NHW participants. Additionally, statin use was found to be a significant factor within the AfAm subset (**Figure S1**). Where the outcome was CVD mortality, age, aspirin use, and BMI were significant factors for the AfAm subset. In the NHW participants, the predictive factors were limited to current smoking and BMI (**Figure S2**).

**Discussion**

In analysis of the comprehensive data on CVD endpoints and covariates from the VITAL trial, we found a striking reduction in MI among AfAm individuals, but not NHW individuals, as a result of randomization to n-3 HUFA supplementation versus placebo. The odds ratio (OR) estimate was 0.17, with a confidence interval (CI) ranging from 0.048 to 0.60. This finding may be of fundamental importance, considering the disproportionately elevated rates of MI incidence and worse outcomes within this demographic [33]. While the VITAL trial's overall analysis did not show a significant reduction in major cardiovascular events across racial groups, our targeted analyses revealed the benefits of n-3 HUFA supplementation may vary for AfAm when compared to NHW. This outcome aligns with previous secondary analyses [16-18] and suggests that aggregating data across racial groups may obscure critical findings related to race-specific effects. Importantly, these data provide evidence that n-3 HUFA supplementation could play a key role in addressing the elevated MI risk among AfAm populations.

Recognizing the initial disparities in covariate distributions between AfAm and NHW participants, our study leveraged propensity score matching to achieve a more balanced and equitable comparison. This strategic approach not only refined the NHW sample for parity but also allowed us to simulate a RCT scenario that compensated for the differences in baseline characteristics. This, in turn, enabled a more reliable assessment of the effect of randomization to n-3 HUFA supplementation versus placebo for AfAm and NHW subgroups. These compelling findings underscore the urgent need for a deeper exploration into whether n-3 HUFA supplementation differentially influences health outcomes across racial (and ethnic) subgroups. A well-designed, sufficiently powered randomized clinical trial focused exclusively on evaluating the hypothesis that n-3 HUFA supplementation has a pronounced effect on reducing MI risk among AfAm population is warranted. Such investigations could yield crucial insights into mechanistic drivers of n-3 supplementation effects by race and inform future precision approaches to cardiovascular care, ultimately facilitating the development of more effective prevention strategies to modify MI risk.

There are known molecular and genetically influenced mechanisms that may explain why n-3 HUFAs, and particularly EPA-enriched supplementation reduces the risk of MI in AfAm but not NHW. Numerous studies have established that the efficiency of the HUFA biosynthetic pathway is under genetic control, with variations in the *FADS* gene cluster exerting the most significant influence on n-6 and n-3 HUFA levels [34]. The link between *FADS* variants and complex lipid and inflammatory phenotypes is well-documented. In fact, the *FADS1-FADS2* gene region has been identified as a critical multimorbidity-associated cluster in the human genome [35]. *FADS* associations have been confirmed in both NHW and AfAm cohorts, with a meta-analysis involving approximately 8,000 AfAm participants validating the connection between *FADS* single nucleotide polymorphisms (SNPs) and lipid phenotypes as well as coronary artery disease [36].

Further, the frequencies of *FADS* genetic variants that affect HUFA biosynthesis, especially ARA, vary significantly across human populations [37]. Although self-reported race does not equate to genetic ancestry, there is a strong correlation in the context of *FADS* variation [13,14,37]. Approximately 80% of AfAm have homozygous 'derived' alleles linked to efficient HUFA metabolism, in contrast to about 43% of NHW individuals.

At the population level, these genetic and other factors significantly alter the balance between the n-6 HUFA, ARA, and n-3 HUFAs, such EPA and DHA [13,14]. Importantly, epidemiology studies highlight a strong association between higher EPA/ARA ratios and a reduced risk of future CVD events [38,39].

If the ratio of ARA to EPA or DHA indeed reflects the synthesis of n-6 and n-3 HUFA-derived oxylipins, as supported by research [40], then AfAm may experience a significant imbalance towards pro-inflammatory and pro-thrombotic versus anti-inflammatory/anti-thrombotic/pro-resolutive oxylipins leading to MI [40,41]. The serious consequences of such an imbalance were underscored two decades ago with the removal of selective cyclooxygenase 2 inhibitors from the market, following cardiovascular events driven by a detrimental shift in the balance between pro-thrombotic thromboxane and anti-thrombotic prostacyclin [42,43]. Collectively, this leads to the hypothesis that the levels and ratios of dietary n-6/n-3 PUFAs-influenced by the Western diet [44] result in a marked imbalance in ARA versus EPA and DHA, along with their oxylipin derivatives. This imbalance is notably exacerbated in AfAm populations where the genetic variation in *FADS* and other race-associated factors substantially enhances the capacity for ARA production [13-15], thereby tipping the balance toward ARA and its downstream pro-inflammatory and prothrombotic oxylipins, and away from EPA and DHA with their anti-inflammatory and anti-thrombotic effects. As a result, supplementation with n-3 HUFAs, which are known to balance the ratio of ARA to EPA and enhance EPA-derived oxylipins [45], could be especially beneficial for AfAm populations, where the extent of this imbalance is so pronounced.

The primary limitation of this study is its post-hoc nature. The VITAL study was not initially designed to examine racial differences in response to n-3 HUFA supplementation versus placebo. Despite efforts to address limitations of post-hoc analyses using propensity score matching, these findings should be interpreted with caution, as they do not directly infer causality. Moreover, while propensity score matching attempts to minimize confounding in observational studies, it relies on the assumption that all relevant variables have been measured and included in the model. This method also reduces sample size, potentially decreasing statistical power and increasing the risk of Type II errors. Despite this, findings from this study underscore the importance of further investigating racial differences in MI risk related to n-3 HUFA supplementation. These findings enhance our understanding of MI health disparities in AfAm and emphasize the need for more focused hypothesis-driven research to establish causal relationships between n-3 HUFA supplementation and health in AfAm.

**Conclusions**

Conflicting results from clinical trials have led to general confusion about the cardioprotective effects of n-3 HUFA supplementation. Our studies suggest that racial/ethnic differences, including genetic variants affecting HUFA metabolism, are likely critical factors that contribute to the variability in clinical trial outcomes, especially for MI. This study also highlights the potential cardioprotective benefits of n-3 HUFA supplementation in AfAm populations and the urgent need for more research to understand the effectiveness of n-3 HUFA supplementation across diverse racial/ethnic populations.

**Author Contribution.** SS, BH, LJ and AH analyzed the data. JCW, CAT, SMS, SS, JU. GY, HHZ and FHC contributed to interpretation of the data. FHC, HHZ and JCW conceptualized and designed the study. CAT, SMS and FHC wrote the manuscript. All authors contributed to critical editing of the manuscript.

**Conflict of interest.** Dr. Chilton is a cofounder of Resonance Pharma, Inc. This company develops diagnostics for lipid targets. This relationship is managed by the Office for Responsible Outside Interests at the University of Arizona. None of the other authors have potential conflicts to report.

**Funding.** This study was funded by NIH (National Center for Complementary and Integrative Health) [R01 AT008621; FHC] and U.S. Department of Agriculture [ARZT-1361680-H23-157; FHC].

**Data availability.** Data from the original trial are available at https://data.projectdatasphere.org/projectdatasphere/html/content/454. The analysis code used in the current study is available at https://github.com/ShudongSun/VITAL_reanalysis/blob/main/PaperCode.R.

**References**

1. GISSI-Prevenzione Investigators. Dietary supplementation with n-3 polyunsaturated fatty acids and vitamin E after myocardial infarction: results of the GISSI-Prevenzione trial. Lancet **1999**, 354, 447-455.

2. Kromhout, D.; Giltay, E. J.; Geleijnse, J. M. n-3 Fatty acids and cardiovascular events after myocardial infarction. N. Engl. J. Med. **2010**, 363, 2015-2026.

3. Rauch, B.; Schiele, R.; Schneider, S.; Diller, F.; Victor, N.; Gohlke, H.; et al. OMEGA, a randomized, placebo-controlled trial to test the effect of highly purified omega-3 fatty acids on top of modern guideline-adjusted therapy after myocardial infarction. Circulation **2010**, 122, 2152-2159.

4. Abdelhamid, A. S.; Brown, T. J.; Brainard, J. S.; Biswas, P.; Thorpe, G. C.; Moore, H. J.; et al. Omega-3 fatty acids for the primary and secondary prevention of cardiovascular disease. Cochrane Database Syst. Rev. **2018**, 11.

5. ASCEND Study Collaborative Group. Effects of n-3 fatty acid supplements in diabetes mellitus. N. Engl. J. Med. **2018**, 379, 1540-1550.

6. Manson, J.E.; Cook, N.R.; Lee, I.M.; Christen, W.; Bassuk, S.S.; Mora, S.; Gibson, H.; Albert, C.M.; Gordon, D.; Copeland, T.; D'Agostino, D. Marine n−3 fatty acids and prevention of cardiovascular disease and cancer. N. Engl. J. Med. **2019**, 380, 23-32.

7. Bhatt, D. L.; Steg, P. G.; Miller, M.; Brinton, E. A.; Jacobson, T. A.; Ketchum, S. B.; et al. Cardiovascular risk reduction with icosapent ethyl for hypertriglyceridemia. N. Engl. J. Med. **2019**, 380, 11-22.

8. Shen, S.; Gong, C.; Jin, K.; Zhou, L.; Xiao, Y.; Ma, L. Omega-3 fatty acid supplementation and coronary heart disease risks: a meta-analysis of randomized controlled clinical trials. Front. Nutr. **2022**, 9, 809311.

9. Hu, Y.; Hu, F. B.; Manson, J. E. Marine omega-3 supplementation and cardiovascular disease: an updated meta-analysis of 13 randomized controlled trials involving 127,477 participants. J. Am. Heart. Assoc. **2019**, 8, e013543.

10. Hoang, T.; Kim, J. Comparative effect of statins and omega-3 supplementation on cardiovascular events: meta-analysis and network meta-analysis of 63 randomized controlled trials including 264,516 participants. Nutrients **2020**, 12, 2218.

11. Elagizi, A.; Lavie, C. J.; O'Keefe, E.; Marshall, K.; O'Keefe, J. H.; Milani, R. V. An update on omega-3 polyunsaturated fatty acids and cardiovascular health. Nutrients **2021**, 13, 204.

12. Zhou, Y.; Khan, H.; Xiao, J.; Cheang, W. S. Effects of arachidonic acid metabolites on cardiovascular health and disease. Int. J. Mol. Sci. **2021**, 22, 12029.

13. Mathias, R. A.; Sergeant, S.; Ruczinski, I.; Torgerson, D. G.; Hugenschmidt, C. E.; Kubala, M.; et al. The impact of FADS genetic variants on ω-6 polyunsaturated fatty acid metabolism in African Americans. BMC Genet. **2011**, 12, 1-0.

14. Sergeant, S.; Hugenschmidt, C. E.; Rudock, M. E.; Ziegler, J. T.; Ivester, P.; Ainsworth, H. C.; et al. Differences in arachidonic acid levels and fatty acid desaturase (FADS) gene variants in African Americans and European Americans with diabetes or the metabolic syndrome. Br. J. Nutr. **2012**, 107, 547-555.

15. Hester, A. G.; Murphy, R. C.; Uhlson, C. J.; Ivester, P.; Lee, T. C.; Sergeant, S.; et al. Relationship between a common variant in the fatty acid desaturase (FADS) cluster and eicosanoid generation in humans. J. Biol. Chem. **2014**, 289, 22482-22489.

16. Manson, J. E.; Bassuk, S. S.; Cook, N. R.; Lee, I. M.; Mora, S.; Albert, C. M.; et al. Vitamin D, marine n-3 fatty acids, and primary prevention of cardiovascular disease current evidence. Circ. Res. **2020**, 126, 112-128.

17. Chilton, F. H.; Manichaikul, A.; Yang, C.; O'Connor, T. D.; Johnstone, L. M.; Blomquist, S.; et al. Interpreting clinical trials with omega-3 supplements in the context of ancestry and FADS genetic variation. Front. Nutr. **2022**, 8, 808054.

18. Bassuk, S.S.; Chandler, P.D.; Buring, J.E.; Manson, J.E.; VITAL Research Group. The VITamin D and OmegA-3 TriaL (VITAL): do results differ by sex or race/ethnicity? Am. J. Lifestyle Med. **2021**, 15, 372-391.

19. Ogata, S.; Manson, J. E.; Kang, J. H.; Buring, J. E.; Lee, I. M.; Nishimura, K.; et al. Marine n-3 fatty acids and prevention of cardiovascular disease: a novel analysis of the VITAL trial using win ratio and hierarchical composite outcomes. Nutrients **2023**, 15, 4235.

20. DeFranco, E. A.; Valentine, C. J.; Carlson, S. E.; Sands, S. A.; Gajewski, B. J. Racial disparity in efficacy of docosahexaenoic acid supplementation for prevention of preterm birth: secondary analysis from a randomized, double-blind trial. Am. J. Obstet. Gynecol. MFM **2024**, 101358.

21. Thoemmes, F. J.; Kim, E. S. A systematic review of propensity score methods in the social sciences. Multivariate Behav. Res. **2011**, 46, 90-118.

22. Zakrison, T. L.; Austin, P. C.; McCredie, V. A. A systematic review of propensity score methods in the acute care surgery literature: avoiding the pitfalls and proposing a set of reporting guidelines. Eur. J. Trauma Emerg. Surg. **2018**, 44, 385-395.

23. Stuart, E. A.; King, G.; Imai, K.; Ho, D. MatchIt: nonparametric preprocessing for parametric causal inference. J. Stat. Softw. **2011**.

24. Hansen, B. B.; Klopfer, S. O. Optimal full matching and related designs via network flows. J. Comput. Graph. Stat. **2006**, 15, 609-627.

25. Gu, X. S.; Rosenbaum, P. R. Comparison of multivariate matching methods: structures, distances, and algorithms. J. Comput. Graph. Stat. **1993**, 2, 405-420.

26. Sekhon, J. S. Multivariate and propensity score matching software with automated balance optimization: the matching package for R. J. Stat. Softw. Forthcoming **2008**.

27. Therneau, T.; Lumley, T. Package Survival: A Package for Survival Analysis in R. R Package 2015, Version 2, 38.

28. Tibshirani, R. Regression shrinkage and selection via the lasso. J. R. Stat. Soc. Series B Stat. Methodol. **1996**, 58, 267-288.

29. Friedman, J.; Hastie, T.; Tibshirani, R. Regularization paths for generalized linear models via coordinate descent. J. Stat. Softw. **2010**, 33, 1.

30. Efron, B.; Tibshirani, R.J. An Introduction to the Bootstrap; Chapman and Hall/CRC: Boca Raton, FL, USA, 1994.

31. Breiman, L. Classification and regression trees; Routledge: Abingdon, Oxfordshire, United Kingdom, 2017.

32. Therneau, T.; Atkinson, B.; Ripley, B.; Ripley, M. B. Package 'rpart'. Available online: https://cran.r-project.org/web/packages/rpart/rpart.pdf (accessed on 17 April 2024), 2015.

33. Office of Minority Health. Heart Disease and African Americans. U.S. Department of Health & Human Services. Available online: https://minorityhealth.hhs.gov/heart-disease-and-african-americans (accessed on 17 April 2024).

34. Chilton, F. H.; Dutta, R.; Reynolds, L. M.; Sergeant, S.; Mathias, R. A.; Seeds, M. C. Precision nutrition and omega-3 polyunsaturated fatty acids: a case for personalized supplementation approaches for the prevention and management of human diseases. Nutrients **2017**, 9, 1165.

35. Fadason, T.; Schierding, W.; Lumley, T.; O'Sullivan, J. M. Chromatin interactions and expression quantitative trait loci reveal genetic drivers of multimorbidities. Nat. Commun. **2018**, 9, 5198.

36. Lettre, G.; Palmer, C. D.; Young, T.; Ejebe, K. G.; Allayee, H.; Benjamin, E. J.; et al. Genome-wide association study of coronary heart disease and its risk factors in 8,090 African Americans: the NHLBI CARe Project. PLoS Genet. **2011**, 7, e1001300.

37. Harris, D. N.; Ruczinski, I.; Yanek, L. R.; Becker, L. C.; Becker, D. M.; Guio, H.; et al. Evolution of hominin polyunsaturated fatty acid metabolism: from Africa to the New World. Genome Biol. Evol. **2019**, 11, 1417-1430.

38. Chiusolo, S.; Bork, C. S.; Gentile, F.; Lundbye-Christensen, S.; Harris, W. S.; Schmidt, E. B.; et al. Adipose tissue n-3/n-6 fatty acids ratios versus n-3 fatty acids fractions as predictors of myocardial infarction. Am. Heart J. **2023**, 262, 38-48.

39. Nelson, J. R.; Raskin, S. The eicosapentaenoic acid: arachidonic acid ratio and its clinical utility in cardiovascular disease. Postgrad. Med. **2019**, 131, 268-277.

40. Nayeem, M. A. Role of oxylipins in cardiovascular diseases. Acta Pharmacol. Sin. **2018**, 39, 1142-1154.

41. Caligiuri, S. P.; Parikh, M.; Stamenkovic, A.; Pierce, G. N.; Aukema, H. M. Dietary modulation of oxylipins in cardiovascular disease and aging. Am. J. Physiol. Heart Circ. Physiol. **2017**, 313, H903-H918.

42. Yedgar, S.; Krimsky, M.; Cohen, Y.; Flower, R. J. Treatment of inflammatory diseases by selective eicosanoid inhibition: a double-edged sword? Trends Pharmacol. Sci. **2007**, 28, 459-464.

43. Bing, R. J.; Lomnicka, M. Why do cyclo-oxygenase-2 inhibitors cause cardiovascular events? J. Am. Coll. Cardiol. **2002**, 39, 521-522.

44. Blasbalg, T. L.; Hibbeln, J. R.; Ramsden, C. E.; Majchrzak, S. F.; Rawlings, R. R. Changes in consumption of omega-3 and omega-6 fatty acids in the United States during the 20th century. Am. J. Clin. Nutr. **2011**, 93, 950-962.

45. Schuchardt, J. P.; Schmidt, S.; Kressel, G.; Willenberg, I.; Hammock, B. D.; Hahn, A.; et al. Modulation of blood oxylipin levels by long-chain omega-3 fatty acid supplementation in hyper-and normolipidemic men. Prostaglandins Leukot. Essent. Fatty Acids **2014**, 90, 27-37.

46. Zhong, M.; Tandon, R. Learning fair classifiers via min–max f-divergence regularization. Proc. 59th Annu. Allerton Conf. Commun. Control Comput. **2023**, IEEE, 1–8.

47. Zhong, M.; Tandon, R. Intrinsic fairness–accuracy tradeoffs under equalized odds. Proc. IEEE Int. Symp. Inf. Theory (ISIT) **2024**, IEEE, 1–6.

48. Zhong, M.; Tandon, R. Splitz: Certifiable robustness via split Lipschitz randomized smoothing. IEEE Trans. Inf. Forensics Secur. **2025**, 20, 1–12.

49. Zhang, W.; et al. Filtered randomized smoothing: A new defense for robust modulation classification. Proc. IEEE Mil. Commun. Conf. (MILCOM) **2024**, IEEE, 1–6.

50. Zhong, M.; Tandon, R. Learning fair robustness via domain mixup. Proc. 58th Asilomar Conf. Signals Syst. Comput. **2024**, IEEE, 1–8.